\documentclass{article}
\usepackage{amsmath}
\usepackage{amssymb}
\usepackage{graphicx}

\begin{document}

\begin{center}
\LARGE \textbf{Physical Reality and Information}
\\[0,25cm]
\LARGE \textbf{- Three Hypotheses}
\normalsize \\[1cm]
Gerd Niestegge\\[0,5cm]
\scriptsize
Zillertalstrasse 39, 81373 M\"{u}nchen, Germany\\
gerd.niestegge@web.de\\
www.gerd-niestegge.de\\[0,5cm]
\normalsize
\end{center}

\normalfont \itshape Abstract. \normalfont Since its emergence, quantum
 mechanics has been a challenge for an understanding of reality which 
 is based on our intuition in a classical world. Nevertheless, it has often 
 been tried to impose this understanding of reality on quantum theory - 
 with limited success. Instead, it might be a better alternative to redefine 
 the meaning of physical reality. This is the objective of the paper. 
 A consideration of the quantum measurement process, conditional 
 probabilities and some well-known typical quantum physical experiments 
 provides the reasoning for the following three hypotheses: (1) Prior 
 to a first measurement, a physical system is not in a quantum state. 
 (2) Physical reality is all that and only that about which (classical) 
 information is available in the universe. (3) Information creation is 
 an independent process and is not covered by the Schr\"{o}dinger 
 equation. It is the first step of the quantum measurement process 
 and does not have a classical counterpart. The first hypothesis 
 makes sense only if the quantum measurement process can be described 
 without presupposing an initial state for the system under consideration. 
 This becomes possible by the objective conditional probabilities 
 which represent the transition probabilities between the outcomes 
 of successive quantum measurements and have been introduced by 
 the author in some recent papers. The second hypothesis holds 
 as well in the classical case, but a certain incompleteness of 
 reality is typical of quantum mechanics and the origin of many 
 quantum phenomena. Classically, the existence of a complete 
 reality is presumed, and hypothesis 3 has no meaning then.\\[0,25cm]

\section{Introduction}

Since its emergence, quantum mechanics has been a challenge for an under\-standing of reality which is based on our intuition in a classical world. Nevertheless, it has often been tried to impose this understanding of reality on quantum theory - with limited success. Instead, it might be a better alternative to redefine the meaning of physical reality.

It is therefore the objective of this paper to make the attempt to propose a definition of what physical reality in a quantum world might be. This will be motivated by the consideration of some well-known typical quantum experiments which shall be discussed at a more abstract level since it is not the intention of the paper to go into the details of their many different technical implementations. The following three hypotheses will then be presented:
\begin{enumerate}
	\item Prior to a first measurement, a physical system is not in a quantum state. 
  \item Physical reality is all that and only that about which (classical) information is available in the universe. 
  \item Information creation is an independent process and is not covered by the Schr\"{o}dinger equation.
\end{enumerate}
Information creation is the first step of the quantum measurement process and does not have a classical counterpart. The second step is the actual perception of the information by an individual observer.
 
The first hypothesis makes sense only if the quantum measurement process can be described without presupposing an initial state for the system under consideration. This becomes possible by the objective conditional probabilities which are the transition probabilities between the outcomes of successive measurements and have been introduced in \cite{ref06} and \cite{ref07}. 

Therefore, the paper starts with a consideration of classical and quantum mechanical conditional probabilities from which the objective conditional probability is derived then. In the subsequent sections, the double-slit experiment and some further experiments are studied to provide the motivation for the second and the third hypothesis.

The second hypothesis holds as well in the classical case, but a certain incompleteness of reality is typical of quantum mechanics and the origin of many quantum phenomena. Classically, the existence of a complete reality is presumed and hypothesis 3 has no meaning then.
  
The paper shall be a rather basic study considering only non-relativistic quantum mechanics, although the three hypotheses might have some impact beyond this. The meaning of information in the hypotheses and in the whole paper is classical since the output of a quantum measurement is classical information. This is emphasized here to avoid any confusion with the so-called quantum information. 

\section{Conditional probability}

Classical probability theory uses a $ \sigma $-algebra of sets as a mathematical model of the system of events, and a probability measure $ \mu $ allocates to each event a number from the unit interval. For a given event $e$ with $\mu(e)>0$, the conditional probability is defined as another probability measure $\mu_e$ satisfying $\mu_e(f)=\mu(f)/\mu(e)$ for all events $f$ with $f \leq e$. The usual notation is to write $\mu(d \mid e)$ instead of $\mu_e(d)$ for any event $d$. A $\sigma$-algebra is a Boolean lattice and satisfies the distributive law $d=d\cap e + d\cap e^c $ which immediately implies that the conditional probability is uniquely defined and has the following shape:
\begin{equation}
\mu(d \mid e) = \mu_e(d) = {\mu(d\cap e)}/{\mu (e)}
\label{eq1}	
\end{equation}
This conditional probability is usually understood as the probability of the event $d$ after the event $e$ has already been observed. The additional knowledge due to the observation of the event $e$ transfers the original probability measure $\mu$ to the new updated probability measure $\mu_e$.

The standard Hilbert space model of quantum mechanics represents the observables as self-adjoint linear operators on the Hilbert space. The observables the spectrum of which contains only the numbers 0 and 1 form the events. Therefore, events are self-adjoint idempotent linear operators. I.e., they are orthogonal projections and there is a one-to-one correspondence between the events and the closed linear subspaces of the Hilbert space. Two specific events are the zero operator 0 and the identity operator $\mathbb{I}$. The event $\mathbb{I} - e$ is also denoted by $e'$ and is interpreted as the negation of the event $e$. A pair of events $e$ and $f$ is called orthogonal if their operator product vanishes (i.e., $ef=0$); the interpretation is that the two events mutually exclude each other. We write $f \leq e$ if the identity $ef=f$ holds; this coincides with the usual order relation for self-adjoint operators and is interpreted as the logical implication.

A probability measure $\mu$ - now also called state - allocates to each event a number from the unit interval and satisfies $\mu(\mathbb{I}) = 1$ and $\mu(e+f)=\mu(e)+\mu(f)$ for any orthogonal event pair $e$ and $f$. Due to Gleason's theorem \cite{ref02} and its generalizations \cite{ref04}, \cite{ref05} such a probability measure has a unique extension to a positive linear functional defined for all bounded linear operators on the Hilbert space (assuming that the dimension of the Hilbert space is not two). This extension is again denoted by $\mu$.

For an event $e$ and a state $\mu$ with $\mu (e) > 0 $, in this situation, a conditional probability shall again be another probability measure $\mu_e$ satisfying $\mu_e(f)=\mu(f)/\mu(e)$ for all events $f$ with $f \leq e$. The issue of conditional probabilities in quantum mechanics has been studied from many different directions (e.g., \cite{ref03}, \cite{ref09}) and \cite{ref08}); here the same simple way of extending the classical conditional probabilities to quantum mechanics is used as, e.g., by Beltrametti and Cassinelli \cite{ref01}.

It shall now be studied whether such conditional probabilities exist and how they look. Assume that $d$ is any further quantum event. The quantum events do not form a Boolean algebra and we cannot use the distributive law. Instead, we have the following decomposition: $d = ede + e'de + ede' + e'de'$. If the conditional probability $\mu_e$ exists, then $\mu_e (e) = 1$, $\mu_e (e') = 0$ and the Cauchy-Schwarz inequality for states implies that $0 = \mu_e (e'de) = \mu_e (ede') = \mu_e (e'de)$; thus $\mu_e (d) = \mu_e (ede)$. Since $ede$ lies in the closed linear hull of those events $f$ with $f\leq e$ by the spectral theorem, we get
\begin{equation}
\mu(d \mid e) = \mu_e (d) = \mu_e (ede) / \mu (e)
\label{eq2}	
\end{equation}
This conditional probability $\mu_e$ is identical with the state that occurs in the L\"{u}ders - von Neumann measurement process: a measurement with the result $e$ transfers the initial state $\mu$ to the final state $\mu_e$. Therefore, the quantum measurement process is identical with the transition from an initial probability measure to the conditional probability where the condition is given by the information provided by the measurement result. In analogy to the classical probabilities, this might suggest to understand a quantum measurement as a mere observation of a certain property of a physical system. However, the assumption that this property already preexists before the measurement takes place is very problematic in quantum mechanics. The conditional probability in equation \ref{eq2} does not behave like a classical one. For instance, it can become independent of the underlying state $\mu$ in some special situations.

\section{Objective probability}

If $ede=\lambda e$ holds for two quantum events $e$ and $d$ with some real number $\lambda$, the conditional probability becomes $ \mu(d \mid e) = \lambda $ and is independent of the state $ \mu$. The value of this probability stems from the algebraic relation between the two events and not from any probability measure or state. It is denoted by $\mathbb{P}(d\mid e)$.

The relation $ede=\lambda e$ holds, for instance, when $e$ is the orthogonal projection on a one-dimensional subspace of the Hilbert space; then $\lambda=\left\langle \psi \mid d \mid \psi \right\rangle / \left\|\psi\right\|^{2}$ with $ \left| \psi \right\rangle$ being any non-zero vector in this subspace, and 
\begin{equation}
\mathbb{P}(d \mid e) = \frac{\left\langle \psi \mid d \mid \psi \right\rangle}{\left\|\psi\right\|^{2}}
\label{eq3}	
\end{equation}
Moreover, if $d$ is the orthogonal projection on another one-dimensional subspace containing the non-zero vector $ \left| \xi \right\rangle$, equation \ref{eq3} becomes
\begin{equation}
\mathbb{P}(d \mid e) = \frac{\mid \left\langle \psi \mid \xi \right\rangle \mid^{2}}{\left\|\psi\right\|^{2}\left\|\xi\right\|^{2}}
\label{eq4}	
\end{equation}
The term on the right-hand side of equation \ref{eq4} is a very familiar quantum mechanical expression and is usually understood as the `transition probability between the states $ \left| \psi \right\rangle$ and $ \left| \xi \right\rangle$', although the interpretation of the square of the absolute value of a complex number as a probability comes a little unmotivated and, furthermore, a transition probability should refer to events and not to states. This interpretation is more or less enforced by the experimental evidence, but not motivated by the mathematical model itself. However, the left-hand side of equation \ref{eq4} has an intrinsic probabilistic interpretation from the very beginning, and it is clear that it is the conditional probability of the event $d$ after the observation of the event $e$ and that this probability does not depend on any initial state of the quantum system.

Outcomes of quantum measurements are events, and $\mathbb{P}(d\mid e)$ is the probability of the outcome $d$ with a future measurement testing $d$ versus $d'$, after a first measurement testing $e$ versus $e'$ has had the outcome $e$. The importance of this special probability $\mathbb{P}(d\mid e)$ lies in the fact that it depends only on the two measurement outcomes ($e$ and $d$), but not on any initial state of the physical system. It is therefore not necessary to assume that a physical system is in a certain, perhaps unknown, quantum state before the measurement testing $e$ versus $e'$ starts. Moreover, $\mathbb{P}(d\mid e)$ is independent of any measuring apparatus or method. 

The special probabilities $\mathbb{P}(\cdot \mid \cdot )$ are transition probabilities between the outcomes of successive measurements and depend on nothing else but the measurement outcomes. If the probability $\mathbb{P}(d\mid e)$ does not exist, the knowledge of the outcome $e$ of the first measurement is not sufficient for any prediction concerning the occurrence of $d$ or $d'$ with a future measurement. However, $\mathbb{P}(d\mid e)$ exists for all events $d$ if $e$ is the orthogonal projection on a one-dimensional subspace of the Hilbert space (i.e., a minimal event) and, in this case, the situation after the measurement can be described by the state $\mu(\cdot):=\mathbb{P}(\cdot\mid e)$.

If the events $e$ and $d$ commute, $ede=ed$ is an event below $e$ and $ede=\lambda e$ can hold only with $\lambda=0$ or $\lambda=1$; i.e., only the trivial cases when $e$ and $d$ are orthogonal or when $e\leq d$ are possible, and either ($\mathbb{P}(d\mid e)=0$ or $\mathbb{P}(d\mid e)=1$. Only these two cases are also possible with classical probabilities. This shows that the non-trivial cases of the special probability $\mathbb{P}(\cdot \mid \cdot )$ are a new non-classical phenomenon. The existence of such non-trivial cases easily follows using equation \ref{eq4} with orthogonal projections $e$ and $d$ on any two different non-orthogonal one-dimensional subspaces of some Hilbert space.

Gathering more information means replacing $e$ by another event $f$ with $f\leq e$ and a minimal non-zero event provides maximum information. In the classical situation, only the two trivial cases with the values 0 and 1  for the conditional probability under a minimal non-zero event occur, but in quantum mechanics non-trivial cases and all values in the unit interval $[0,1]$ are possible. 

So far, it has been seen that the special probability $\mathbb{P}(d\mid e)$  is a property of the event pair $e$ and $d$. It does not depend on any underlying initial state or probability measure nor does it depend on any measuring apparatus or method. It cannot be improved by gathering additional information when the measurement result $e$ is a minimal event (projection on a one-dimensional subspace of the Hilbert space). This shows that the probability $\mathbb{P}(\cdot \mid \cdot )$ has a certain objective character and it shall therefore be called an \emph{objective conditional probability} in this paper.

The significance of objective probability for quantum mechanics has already been recognized by other authors. To base the interpretation of quantum mechanics on the interpretation of objective probability is the last one of Mermin's six desiderata \cite{ref11} and later in this article he writes `Central ... is the doctrine that the only proper subjects of physics are correlations among different parts of the physical world. Correlations are fundamental, irreducible and objective'. Bub and Pitowski \cite{ref10} write `Hilbert space imposes ... objective probabilistic constraints on correlations between events'. The definition of the objective conditional probability via its state-independence provides it with a mathematical foundation as well as a clear interpretation.

\section{The first hypothesis}

The objective conditional probabilities perfectly describe the transition probabilities between measurements without requiring the assumption that the physical system under consideration is in a certain state prior to the first measurement. These probabilities can be tested by physical experiments although the estimation of a probability requires more than one single measurement.

There is neither a stringent reason for the assumption that a physical system is in a certain quantum state prior to a first measurement nor for the assumption that it is not. However, the potential state prior to a first measurement cannot be verified by any physical experiment and the assumption that there is no such state is more appropriate for an empirical science dealing only with experimentally verifiable phenomena. This becomes the first hypothesis. 
\\[0,5cm]
\bfseries \sffamily Hypothesis 1: \normalfont \itshape Prior to a first measurement, a physical system is not in a quantum state. \normalfont
\\[0,5cm]
When a measurement outcome $e$ is a minimal event, the situation after the measurement can be described by the state $\mu(\cdot):=\mathbb{P}(\cdot\mid e)$. Such a first measurement is often called `preparation', although there is no real reason for the distinction between `measurement' and `preparation'. A situation with a `preparation' and a `measurement' is the same as a situation with a `first measurement' and a `second successive measurement'. To `prepare' a specific quantum state, the first measurement must be repeated with new inputs until the desired result is achieved.

Note that the objective conditional probability $\mathbb{P}(d\mid e)$ covers also cases when $e$ is not a minimal event; $\mathbb{P}(d\mid e)$ does then not exist for all $d$, but for some $d$. E.g., consider the two matrices
\begin{center}
$ e= \left(
\begin{array}{cccc}
  1 & 0 & 0 & 0\\
  0 & 1 & 0 & 0\\
  0 & 0 & 0 & 0\\
  0 & 0 & 0 & 0\\
\end{array}
\right)
$
and 
$ d= \frac{1}{2} \left(
\begin{array}{cccc}
  1 & 0 & 1 & 0\\
  0 & 1 & 0 & 1\\
  1 & 0 & 1 & 0\\
  0 & 1 & 0 & 1\\
\end{array}
\right)
$
\end{center}
\mbox{} \\
Then $ede=e/2$ and $\mathbb{P}(d\mid e) = 1/2$ although $e$ is not minimal (it is a projection on a two-dimensional subspace). In this case, the situation after a measurement with the outcome $e$ cannot be described by a quantum state (neither by a pure one nor by a mixed one; the calculation of a mixed state would require the knowledge of the state prior to the measurement). Therefore, the objective conditional probabilities cover more situations than the common `preparation and measurement' approach, where it is assumed that the `preparation' determines a unique state.

Though the proposal to understand the L\"{u}ders - von Neumann measurement as a non-classical probability conditionalization rule has been known for some time \cite{ref03}, the state-independence of the conditional probabilities in certain cases appears to have gained only little attention so far. A deeper understanding of these objective conditional probabilities requires the consideration of repeated conditionalization.

\section{Repeated conditionalization}

The conditional probability of a further event $d$ in the state $\mu$ after having observed a sequence of $n$ events $e_1, e_2, ..., e_n (n>1)$ is inductively defined via
$$ \mu _{e_1,e_2,...,e_n} := \left(\mu _{e_1,e_2,...,e_{n-1}}\right)_{e_n} $$ 
if $\mu _{e_1,e_2,...,e_{n-1}}(e_n)>0$. Again $\mu(d \mid e_1,e_2,...,e_n)$ shall also be written for $\mu _{e_1,e_2,...,e_n}(d)$. With classical probabilities, this becomes
\begin{equation}
\label{eq5}	
\mu(d \mid e_1,e_2,...,e_n) = \frac{\mu(d\cap e_1 \cap \cdots \cap e_n)}{\mu(e_1 \cap \cdots \cap e_n)}	= \mu(d \mid e_1 \cap e_2 \cdots \cap e_n) 
\end{equation}
With the quantum model, it becomes
\begin{equation}
\label{eq6}
\mu(d \mid e_1,e_2,...,e_n) = \frac {\mu (e_1 e_2 \cdots e_n d e_n \cdots e_2 e_1)}{\mu (e_1 e_2 \cdots e_n \cdots e_2 e_1)}	
\end{equation}
If $e_1 e_2 \cdots e_n d e_n \cdots e_2 e_1 = \lambda e_1 e_2 \cdots e_n \cdots e_2 e_1$ for some real number $\lambda$, this conditional probability is again independent of the state and is denoted by $\mathbb{P}(d \mid e_1,e_2,...,e_n)$. This objective conditional probability exists e.g., if the operator product $e_1 e_2 \cdots e_n$ does not vanish and if one of the events is the projection on a one-dimensional subspace. Assume that this is $e_k$ ($1\leq k \leq n $). Then $e_1 e_2 \cdots e_n d e_n \cdots e_2 e_1 = \alpha e_1 e_2 \cdots e_k \cdots e_2 e_1$ and $e_1 e_2 \cdots e_n \cdots e_2 e_1 = \beta e_1 e_2 \cdots e_k \cdots e_2 e_1 $  for some $\alpha\geq 0$ and $\beta > 0$ such that $\mathbb{P}(d \mid e_1,e_2,...,e_n) = \alpha/\beta$.

If $\mathbb{P}(d \mid e_n)$ exists, the objective conditional probability under the event sequence $e_1, e_2, ..., e_n$ exists as well and $\mathbb{P}(d \mid e_1,e_2,...,e_n) = \mathbb{P}(d \mid e_n)$. I.e., the previous observations $e_1, e_2, ..., e_{n-1}$ can completely be ignored in this case.

Moreover, $\mathbb{P}(e_n \mid e_1,e_2,...,e_n) = 1$ always holds, but $\mathbb{P}(e_k \mid e_1,e_2,...,e_n) = 1$ need not equal 1 for $k<n$ (e.g., if $\mathbb{P}(e_k \mid e_n)$ exists and $\mathbb{P}(e_k \mid e_n)\neq 1$). This means that, if the same property is tested a second time without other tests in between, the second test will always provide the same outcome as the first one. However, if other properties have been tested in between, there is a chance that the last test provides another outcome although the same system property is tested again as in the first test. The information gained from the first test seems to have been destroyed by the information about the other properties tested in between.

This behavior of the objective conditional probabilities might be quite surprising from a classical point of view, but is totally in line with quantum experiments (e.g., consider a series of spin measurements along different spatial axes with an electron or photon).

The classical conditional probability $\mu(e_n \mid e_1,e_2,...,e_n)$ does not depend on the sequential order of the events $e_1, e_2, \cdots, e_n$ and is identical with the conditional probability under the single event $e_1 \cap e_2 \cdots \cap e_n$. However, in the quantum case, a logical `and'-operation like $\cap$ is not generally available and observing an event $e_1$ first and an event $e_2$ second becomes different from observing $e_2$ first and $e_1$ second. Timely order seems to have more significance then than in the classical case.

With the common Hilbert space model, the quantum events form a lattice. However, the observation of the event series $e_1, e_2, ..., e_n$ is not identical with the observation of the single event $e_1 \wedge e_2 \cdots \wedge e_n$. E.g., consider two non-orthogonal one-dimensional subspaces of a Hilbert space and the corresponding projection operators $e_1$ and $e_2$; then $e_1 \wedge e_2 = 0$, but $\mathbb{P}(d \mid e_1,e_2) = \mathbb{P}(d \mid e_2) $ as well as $\mathbb{P}(d \mid e_2,e_1) = \mathbb{P}(d \mid e_1) $ both exist for all events $d$. If a logical `and'-operation were available, one would expect `$e_1$ and $e_2$' = $e_2$ as well as `$e_2$ and $e_1$' = $e_1$. This shows that the lattice operation $\wedge$ cannot be a candidate for the logical `and'-operation and, moreover, that a logical `and'-operation cannot make any sense in the quantum case. Only if two events $e_1$ and $e_2$ commute, $e_1 e_2 = e_2 e_1 = e_1 \wedge e_2$ can be considered to be something like `$e_1$ and $e_2$.' 

\section{The double-slit experiment}

Now assume that the event $e$ is the sum of two orthogonal events $e_1$ and $e_2$ and consider the conditional probability of another event $d$ under $e$ in a state $\mu$ with $\mu(e_1)>0$ and $\mu(e_2)>0$. In the classical case, the distributive law again implies the following identity:
\begin{equation}
\label{eq7}
\mu(d \mid e)= \frac{1}{\mu (e)} \left(\mu(d \mid e_1) \, \mu(e_1) + \mu(d \mid e_2) \, \mu(e_2) \right)
\end{equation}
In the quantum case,  the identity $ede=e_1 d e_1 + e_2 d e_2 + e_1 d e_2 + e_2 d e_1$ holds and thus
\begin{equation}
\label{eq8}
\mu(d \mid e)= \frac{1}{\mu (e)} \left(\mu(d \mid e_1) \, \mu(e_1) + \mu(d \mid e_2) \, \mu(e_2) + 2 Re\, \mu (e_1 d e_2) \right)
\end{equation}
where $\,Re\, \mu (e_1 d e_2)$ denotes the real part of a complex number $\mu (e_1 d e_2)$. If neither $e_1$ nor $e_2$ commutes with $d$, the last term $2 Re\, \mu (e_1 d e_2) $ on the right-hand side of equation \ref{eq8} need not vanish (although $e_1$ and $e_2$ are orthogonal) and, moreover, can be negative as well as positive. This term is responsible for a certain deviation from the sum of the first two terms on the right-hand side of equation \ref{eq8} which are identical with the classical case in equation \ref{eq7}. In quantum mechanics, this deviation is called interference and is often explained by allocating wave-like properties to quantum particles.

In the same way, interference occurs with the objective conditional probability $\mathbb{P}(d \mid f,e)$; it is assumed that $f$ is the orthogonal projection on a one-dimensional subspace such that this probability exists. Equation \ref{eq8} then becomes:
\begin{equation}
\label{eq9}
\mathbb{P}(d \mid f,e)= \frac{1}{\mathbb{P}(e \mid f)} \left(\mathbb{P}(d \mid f, e_1) \, \mathbb{P}(e_1 \mid f) + \mathbb{P}(d \mid f, e_2) \, \mathbb{P}(e_2 \mid f) + 2 Re\, \lambda \right)
\end{equation}
where $\,\lambda$ is the complex number with $f e_1 d e_2 f = \lambda f$ and $Re\, \lambda$ its real part.

Now consider the well-known double-slit experiment with a micro-physical particle (e.g., a photon or electron). Let $f$ be the event that the particle has the linear momentum $\vec{p}$. Let $e_k$ $(k=1,2)$ be the events that the particle passes through slit 1 and 2, respectively, and let $d$ be the event that the particle is detected at a certain fixed location $x$ behind the screen with the two slits. Then the different objective conditional probabilities in equation \ref{eq9} own the following interpretations:

\small
\begin{center}
\begin{tabular}{rcl}
$\mathbb{P}(d \mid f,e_1)$ & = & probability that a particle with the linear momentum $\vec{p}$ will\\
 & & be detected at $x$, when slit 1 is open and slit 2 is shut.\\
$\mathbb{P}(d \mid f,e_2)$ & = & probability that a particle with the linear momentum $\vec{p}$ will\\
 & & be detected at $x$, when slit 1 is shut and slit 2 is open.\\
$\mathbb{P}(d \mid f,e)$ & = & probability that a particle with the linear momentum $\vec{p}$ will\\
 & & be detected at $x$, when both slits are open.\\
$\mathbb{P}(e_1 \mid f)$ & = & probability that a particle with the linear momentum $\vec{p}$ will\\
 & & pass through slit 1.\\
$\mathbb{P}(e_2 \mid f)$ & = & probability that a particle with the linear momentum $\vec{p}$ will\\
 & & pass through slit 2.\\
$\mathbb{P}(e \mid f)$ & = & probability that a particle with the linear momentum $\vec{p}$ will\\
 & & pass through any one of the two slits.
\end{tabular}
\end{center}
\normalsize

The interference patterns that are observed with the double-slit experiments with micro-physical particles and that contradict the behavior of the classical conditional probabilities find now their explanation in the interference term $2 Re \, \lambda $ in equation \ref{eq9}. With the equations \ref{eq8} and \ref{eq9}, interference becomes an intrinsic property of the conditional probabilities. Its origin lies in the algebraic structure of the system of quantum events which does not anymore satisfy the distributive law of Boolean algebra.

As soon as it it possible to find out through which one of the two slits the particle passes, however, there is no interference, and instead of equation \ref{eq8} the equation \ref{eq7} or instead of equation \ref{eq9} the following one has to be used:
\begin{equation}
\label{eq10}
\mathbb{P}(d \mid f,e)= \frac{1}{\mathbb{P}(e \mid f)} \left(\mathbb{P}(d \mid f, e_1) \, \mathbb{P}(e_1 \mid f) + \mathbb{P}(d \mid f, e_2) \, \mathbb{P}(e_2 \mid f) \right)
\end{equation}
The fact that some information is available in principle makes an important difference in quantum mechanics. It is not necessary that a human observer knows this information; it is sufficient that it exists. In this cases, the equations \ref{eq7} or \ref{eq10} are valid. When the information exists and is perceived by an observer, the probability valid for this observer becomes the conditional probability $\mu(d \mid e_k)$ or $\mathbb{P}(e_k \mid f)$, assuming that the event $e_k$ has occurred ($k=1$ or $k=2$). 

In quantum mechanics, a two-step process is encountered; the first one is the creation of the information, and the second one is the actual perception of the information by an individual observer. The second step is identical with the transition to a conditional probability in classical probability theory. The first one is specific to quantum mechanics and does not occur with classical probabilities since the mathematical model ($\sigma$-algebras) in principle assumes the availability of all information from the beginning. These considerations shall be continued studying some other experiments.

\section{Further experiments}

A micro-physical particle is sent through a serial arrangement of three measurement apparatuses. Each one tests a certain property (A or B) of the particle; the first one and the last one test the same property (A: $e$ versus $e'$) while the second one in the middle tests another property (B: $f$ versus $f'$). After passing through the first apparatus the particle is blocked if the measurement outcome in this first apparatus is $e'$ and the particle is sent to the second apparatus in the middle only if the outcome is $e$.

A concrete realization of this arrangement can be implemented by using electrons and measuring their spin along the x-axis in the first and in the third apparatus and measuring their spin along the y-axis in the second apparatus in the middle. The event $e$ corresponds to `the spin along the x-axis is $+\hbar/2$',  $e'$ to `the spin along the x-axis is $-\hbar/2$', $f$ to `the spin along the y-axis is $+\hbar/2$', and $f'$ to `the spin along the y-axis is $-\hbar/2$'.

\begin{figure}[h!]
\centering
\includegraphics[width=10cm]{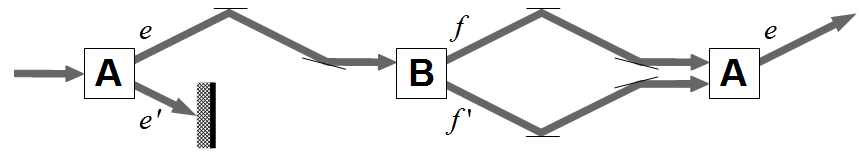}
\caption{The basic arrangement}
\label{fig:Fig1}
\mbox{}\\[1cm]
\includegraphics[width=10cm]{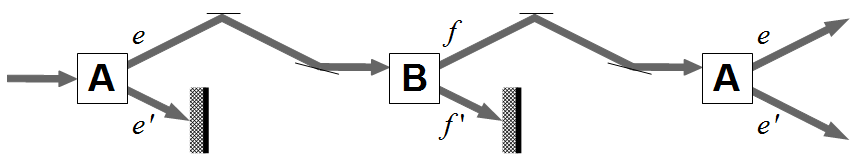}
\caption{Blocking one path}
\label{fig:Fig2}
\mbox{}\\[1cm]
\includegraphics[width=10cm]{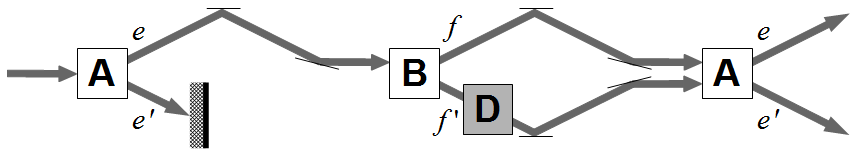}
\caption{Insertion of a detector}
\label{fig:Fig3}
\end{figure}

The basic arrangement is shown in Figure \ref{fig:Fig1}. The two outlet paths of the apparatus in the middle are joined and fed into the third apparatus which tests the same particle property (A) as the first apparatus. Therefore the measurement outcome of the third apparatus is $e$ with probability 1:
\begin{equation}
\label{eq11}
\mathbb{P}(e \mid e,f+f')= \mathbb{P}(e \mid e,\mathbb{I})= \mathbb{P}(e \mid e) = 1
\end{equation}

In the next arrangement (Figure \ref{fig:Fig2}), one of the two outlet paths of the apparatus in the middle is blocked and, surprisingly, the measurement outcome $e'$ becomes possible with a non-zero probability in the third apparatus although this is the negation of $e$ which was the measurement outcome of the first apparatus in the series:
\begin{equation}
\label{eq12}
\mathbb{P}(e \mid e,f)= \mathbb{P}(e \mid f)= 1/2
\end{equation}
Here, the first =-sign holds whenever $f$ is a minimal event and the second one holds in the concrete case with the electron spin. Then $\mathbb{P}(e' \mid e,f) =$ $1 - \mathbb{P}(e \mid e,f) = 1/2$.

In the third arrangement (Figure \ref{fig:Fig3}), a detector D is inserted in one of the outlet paths of the apparatus in the middle. It detects the particle, stores this information, and lets the particle pass. Both outlet paths of the apparatus in the middle are then joined and fed into the third apparatus as in the arrangement shown in Figure \ref{fig:Fig1}. Nevertheless, both measurement outcomes ($e$ and $e'$) do occur in the third apparatus with non-zero probability as they do in the second arrangement (Figure \ref{fig:Fig2}):

\begin{equation}
\label{eq13}
\mathbb{P}(e \mid e, f) \, \mathbb{P}(f \mid e) + \mathbb{P}(e \mid e, f') \, \mathbb{P}(f' \mid e) 
\end{equation}
\mbox{} \hspace{5cm} $ = \mathbb{P}(e \mid f) \, \mathbb{P}(f \mid e) + \mathbb{P}(e \mid f') \, \mathbb{P}(f' \mid e) $\\[0,3cm]
\mbox{} \hspace{5cm} $ = 1/2 $\\[0,25cm]
Here, the first =-sign holds whenever $f$ and $f'$ are minimal events and the second one holds in the concrete case with the electron spin since each of the four conditional probabilities equals $1/2$ then. The probability of the event $e'$ is $\mathbb{P}(e' \mid e,f) = 1 - \mathbb{P}(e \mid e,f) = 1/2$. When the observer reads the information stored in the detector, the probability becomes $\mathbb{P}(e \mid e, f) = \mathbb{P}(e \mid f)$ or $\mathbb{P}(e \mid e, f') = \mathbb{P}(e \mid f')$ depending on whether the particle was detected by D or not. 

When the outcome of the B measurement is known by the observer, the same conditional probability is obtained as in equation \ref{eq12}. Interesting are the difference between the first arrangement displayed in Figure \ref{fig:Fig1} and the third one displayed in Figure \ref{fig:Fig3} when the observer has not checked the information stored in the detector D and the question why different rules for the calculation of the probability have to be applied here.

If a particle uses the upper path in Figure \ref{fig:Fig3}, there is no interaction of it with the detector and nevertheless the presence of the detector in the other path is responsible for a dramatical change of the probabilities for the measurement outcomes in the third apparatus. If the detector is present, the probability of $e'$ is non-zero while it is zero without the detector (Figure \ref{fig:Fig3}). Only the mere fact that information about the path of the particle is available in principle can be the reason for the different probabilities; then equation \ref{eq13} is the rule for the calculation of the probability while this is equation \ref{eq11} in the case when no such information exists.

Instead of using electrons, similar experiments can be executed with with photons and their spin (i.e., with light and its polarization). Even small atoms have been used in such experiments and particularly in the third arrangement (Figure \ref{fig:Fig1}), where an excited atom then emits a photon in the detector D rendering possible the later look-up whether or not the atom has passed through the detector. 

\section{Physical reality and the second hypothesis}

The experiments considered in the last two sections seem to indicate that a physical phenomenon `becomes reality' in some cases and that it does not in other ones. Depending on this, different rules for the probability calculation have to be used. With the arrangement of Figure \ref{fig:Fig1}, none of the two paths between the second and the third apparatus `becomes reality' while the path does `become reality' in the other two cases (Figures \ref{fig:Fig2} and \ref{fig:Fig3}). Also with the double-slit experiment, it does neither `become reality' that the particle passes through slit one nor that is passes through slit two; `reality becomes' only that it passes through one of them without specifying which one. First, it must be noted that physical reality is meant here; philosophy may consider other non-physical realities. Second, the wording `to become reality' does not have a clear meaning as long as there is no definition of what physical reality is. To find such a definition is the objective of this section. 

The creation of reality appears to be identical with the creation of information. This suggests that something is physically real if information about it is available.

On the other hand, physical reality must be experimentally verifiable - at least in principle.  This requires the availability of information stored in nature somehow, and something about which no information is available cannot be considered a part of physical reality. If something is physically real, information about it must be available. These considerations are summarized in the following hypothesis as definition of physical reality.
\\[0,5cm]
\bfseries \sffamily Hypothesis 2: \normalfont \itshape Physical reality is all that and only that about which some information is stored somewhere somehow in the universe.\normalfont
\\[0,5cm]
This does not require a (human) observer; relevant is only the fact that the information is stored such that it is available to a potential observer. The meaning of information is classical here.

Moreover, a measuring apparatus set up by a human is not necessarily required. The same physical phenomenon that happens in such an apparatus also occurs in nature without a human being involved. An apparatus is nothing else but a part of nature intentionally arranged by a human to study a particular phenomenon.

The typical quantum phenomena require that something does not become reality or that no information about it is created. With the double-slit experiment, this is the actual path through either the first or the second slit. With the arrangement shown in Figure \ref{fig:Fig1}, this is the measurement outcome in the apparatus in the middle or the actual path between the this one and the third apparatus.

Hypothesis 2 itself is also valid in a classical theory. It is not this hypothesis that distinguishes the quantum case from the classical one, but the incompleteness of the reality. Sometimes, quantum events do not become reality; no information exists in the universe whether they are true or false. In the classical case, it is always presumed that each event is either true or false, and probabilities arise from a lack of knowledge about the reality. In the quantum case, however, it is well-known that it is impossible to allocate a true- or false-value in a consistent manner to each event \cite{ref13}; this means that a complete reality is not possible.

\section{Quantum measurement and the third hypo\-thesis}

In quantum mechanics, the Schr\"{o}dinger equation results in unitary time evolution operators. The time evolution can be allocated to the states, which is the Schr\"{o}dinger picture,  or to the observables, which is the Heisenberg picture. Mathematically, the two pictures are equivalent. However, in view of the first hypothesis, the Heisenberg picture is to be preferred. One of the questions around the quantum measurement process is whether this type of time evolution is universally valid such that is also covers the quantum measurement process. In the wording of the previous section, this question becomes whether the creation of information is coverd by the Schr\"{o}dinger equation. 

\begin{figure}[h!]
\centering
\includegraphics[width=10cm]{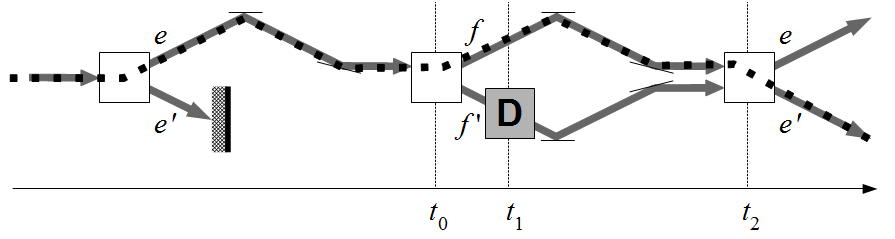}
\caption{The arrangement of Figure 3 with time scale}
\label{fig:Fig4}
\end{figure}

To study this, the experiment depicted in Figure \ref{fig:Fig3} shall be reconsidered. Suppose that the measurement outcome for a certain particle is $f$ in the apparatus in the middle and is $e'$ in the third apparatus. Such a path is shown in Figure \ref{fig:Fig4} where a time scale has also been added. The question at which time the creation of information happens shall now be addressed.
\newpage
At what time $t$ is the information created that the outcome in the apparatus in the middle is $f$? This is not the time $t=t_0$ when the particle passes through the apparatus in the middle, but the time $t=t_1$ when it would have passed through the detector in the case of the outcome $f'$. Until $t=t_1$ the detector could still be removed and this information would never be created as in the arrangement depicted in Figure \ref{fig:Fig1}. At $t=t_1$ the information is created that the measurement outcome at the earlier time $t=t_0$ was $f$; at $t=t_0$ itself this information was not yet available. 

Moreover, the fact that the particle does not pass through the detector implies that it owns the property $f$ in the time interval $[t_0,t_2]$ on its path between the second and the third apparatus. So, at time $t=t_1$, an information is created at the detector D that the particle owns the property $f$ at this time point although the particle is located at another distant place at this time point. Therefore, it is possible to create information about a certain phenomenon happening spatially and timely separated from the location where the information is generated. This motivates the last hypothesis.
\\[0,5cm]
\bfseries \sffamily Hypothesis 3: \normalfont \itshape The process which creates information or physical reality is an independent process and not covered by the unitary time evolution following from the Schr\"{o}dinger equation.\normalfont
\\[0,5cm]
Classically, the existence of a complete reality or the availability of complete information is always presumed and hypothesis 3 has no meaning then. Information creation is the first step of the quantum measurement process and has no counterpart in classical physics. The second step is then the actual perception of the information by an individual observer; as in the classical case, it is a mere observation of a preexisting reality since that what is observed has been created in the first step.

In daily life it is quite common that that information is deleted. Information written on a piece of paper is destroyed by shredding or burning the paper. The bits stored in computer are deleted by overwriting them with other bits. If no copy of the information exits on another piece of paper, another computer or in the brain of a human or in any other form, only then the information is really destroyed. However, as long as copies exist, the information is still there, and even if no copies exist, it may still be possible to reconstruct the information by some physical process. 

With the Schr\"{o}dinger equation, the complete past and future evolution of a state can determined from the state at a certain time point. The belief in the universal validity of the Schr\"{o}dinger equation would include that information about a physical system at one point in time should determine it at any other time and that physical information cannot be destroyed therefore. 
\newpage
With hypothesis 3, however, there is no reason anymore to rule out that information can be deleted. In view of hypothesis 2, this gets a higher significance. \emph{If} the deletion of information were possible, this might have important consequences in physics and philosophy which are beyond the scope of the present paper. 

\section{Conclusions}

An understanding of reality based on our intuition in a classical world is not anymore appropriate for the quantum world, and it has therefore been proposed to redefine the meaning of physical reality.

It has been argued that only the measurement outcomes represent physical reality. It is not necessary that a measurement outcome is perceived by a human observer; it is sufficient that some information about the outcome is available to a potential observer. Moreover, a measuring apparatus set up by a human is not necessarily required. The same physical phenomena that happen in such an apparatus also occur in nature without a human being involved. An apparatus is nothing else but a part of nature intentionally arranged by a human to study a particular phenomenon.

Thus physical reality becomes identical with the information available in the universe about the natural phenomena. The meaning of information is classical here. A major difference to the classical case is the incompleteness of reality. This  incompleteness is a major source for the typical quantum phenomena. Quantum interference occurs only if certain events do not become reality.

Moreover, the creation of information or, in other words, the quantum measurement process, is an independent process. It is not covered by the unitary time evolution following from the Schr\"{o}dinger equation. There is no common agreement on this view, but some other authors who have recently supported it are J. Bub \cite{ref10}, I. Pitowsky \cite{ref12} and J. Rau \cite{ref14}.

Since only the measurement outcomes represent physical reality, a quantum measurement should not be described as a transition between states, but as a transition between measurement outcomes. Since this transition is not deterministic, transition probabilities between the measurement outcomes are required. These are the objective conditional probabilities considered in sections 3 and 4 of this paper.

\end{document}